\documentclass{article}
\usepackage{graphicx}
\usepackage[T1]{fontenc}
\usepackage{authblk}

\title{Bell-type Polarization Experiment With Pairs Of Uncorrelated Optical Photons}
\author[1]{M. Iannuzzi}
\author[2]{R. Francini   \thanks{roberto.francini@roma2.infn.it}}
\author[1]{R. Messi      \thanks{roberto.messi@roma2.infn.it}}
\author[3]{D. Moricciani \thanks{dario.moricciani@lnf.infn.it}}
\affil[1]{Phys. Dept. Universty of Roma "Tor Vergata", Via della Ricerca Scientifica 1, 00133, Rome}
\affil[2]{Dept. of Industrial Engineering, University of Rome "Tor Vergata", Via del Politecnico, 1, 00133, Rome}
\affil[3]{INFN - LNF, Via E. Fermi 54, 00044 Frascati}

\begin{document}
\maketitle

\begin{abstract}
We present a Bell-type polarization experiment using two independent
sources of polarized optical photons, and detecting the temporal
coincidence of pairs of uncorrelated photons which have never been
entangled in the apparatus. Very simply, our measurements have tested
the quantum-mechanical equivalent of the classical Malus' law on an
incoherent beam of polarized photons obtained from two separate and
independent laser sources greatly reduced in intensities.The outcome
of the experiment gives evidence of violation of the Bell-like
inequalities. Drawing the conclusions of the present work, we invoke
the distinction between the concepts of state-preparation and 
measurement to understand this result.
\end{abstract}

\section{Introduction and Motivation for the Experiment} 
The Bell theorem \cite{ref1} is a theorem aimed at proving that, for a certain family 
of experiments, the outputs required to satisfy any local theory are in contradiction 
with those predicted by the quantum formalism. Such experiments have actually been 
performed involving pairs of entangled particles, generally pairs of 
polarization-correlated optical photons \cite{ref2,ref3,ref4,ref5} or spin-1/2 massive  
particles in the singlet state \cite{ref6}. Violation of the Bell inequality, clearly 
observed since the early measurements and giving evidence in favour of the quantum 
theory, has more recently confirmed definitively in three tests \cite{ref7,ref8,ref9}  
performed free of the "detection loophole" and the "locality loophole" i.e. the two 
experimental loopholes that before had never been closed simultaneously and unquestionably 
in a single experiment, and in another experiment closing also the "freedom-of-choice 
loophole" \cite{ref10}. The mentioned loopholes are clearly defined in the 
introductions of these papers. Our motivation to perform a Bell-type polarization 
experiment with pairs of uncorrelated optical photons emitted from two independent 
sources and never entangled in the apparatus, was the following reflection. 
Let us first remind the conceptual scheme of a Bell-type experiment. Consider the 
two-component physical system formed of two entangled particles, and two instruments 
(consisting of polarizer and detector), each associated with one component. The 
detectors record counts of the arriving particles. Each instrument has a range of 
settings, whose possible values are denoted as {\bf a} for the first instrument 
and {\bf b} for the second 
instrument. 
The physical quantities that will be measured are the joint probabilities of 
simultaneous detection of the two particles. The results of a measurement depend 
also on uncontrolled parameters $\lambda$ (hidden variables). A Bell-like inequality, 
like the ones derived by CHSH \cite{ref11} or by CH \cite{ref12}, is always satisfied 
by physical systems formed of pairs of particles whose components are spatially separated 
and independent of each other (locality). In particular, the basic mathematical form 
of the CH inequality, upon averaging over the  $\lambda$-distribution, is:

\begin{equation}
\frac{p_{12}({\bf a}, {\bf b}) - p_{12}({\bf a}, {\bf b}') + p_{12}({\bf a}', {\bf b}) + 
p_{12}({\bf a}', {\bf b}')}{p_1({\bf a'}) + p_2({\bf b})} \le 1
\label{eq1}
\end{equation}

\noindent where $p_{12}({\bf a}, {\bf b}) = \int P_{12}({\bf a}, {\bf b}, 
\lambda)\rho(\lambda)d\lambda$ is the probability of detecting both components of 
the pair, $p_1({\bf a}) = \int P_1({\bf a}, \lambda) \rho(\lambda)d\lambda$ and 
$p_2({\bf b}) = \int P_2({\bf b}, \lambda) \rho(\lambda)d\lambda$ are respectively 
the single-count probabilities of detector \#1 and detector \#2 recording a count, 
and $\rho(\lambda)$ is the probability distribution of the uncontrolled parameters 
$\lambda$.
Results (\ref{eq1}) has been obtained by assuming $P_{12}({\bf a}, {\bf b}, \lambda) 
= P_1({\bf a}, \lambda) P_2({\bf b}, \lambda)$, i.e. by assuming that there is 
no action at a distance between instrument \#1 and instrument \#2 (locality). Note that, 
describing our experiment in the following, we shall show that the conditions of 
separability and independence are experimentally satisfied in our measurements.

Let us now emphasize that the derivation of Bell-like inequalities, among which the CH 
inequality,  "make no use of quantum mechanics but only some much simpler postulates, 
among which Einstein’s locality principle is the most prominente" \cite{ref13}\footnote{
In this modern text-book, the issue is presented very clearly.}
As recalled above, 
for a quantum system formed of two polarization-entangled components whose
correlation function $p_{12}({\bf a}, {\bf b}) = 1/2\cos^2(\vartheta)$ depends 
only on the relative angle $\vartheta$ between the orientations of the 
polarization-measuring instruments, violation of (\ref{eq1}) has been 
measured for a wide range of $\vartheta$.  
Let us consider the joint probability $p_{12}({\bf a}, {\bf b}) = p_1({\bf a}, 1) 
\cdot p_2({\bf b}, 2|1)$, where $p_1({\bf a}, 1)$ is the probability of the 
photon \#1 passing through the filter oriented in the direction {\bf a}, and 
$p_2({\bf b}, 2|1)$ is the conditional probability of photon \#2 passing 
through the filter oriented in the direction {\bf b} if the first photon 
passes.  Since the derivation of (\ref{eq1}) does not make use of quantum 
mechanics and the joint probability contains information about both measured 
particles, we cannot exclude the possibility that it may show correlation also 
with a physical system formed of two not-entangled particles. Consequently, we 
cannot exclude that, if for particular experimental configurations such a correlation 
depends only on he relative angle $\vartheta$, violation of Bell's theorem may 
occur, similarly to the quantum case.
Notice that Bell-type polarization experiments using two independent 
photon sources have already been performed, but, differently from the present work,
with the two photons of a pair which have been entagled in the apparatus \cite{ref14,ref15}. 
Actually we have found that an experiment aimed at testing our reflection 
was feasible, and we have performed it. 

Generally, in the experiments performed up to date to test the predictions of Bell's 
theorem, the parameters {\bf a} and {\bf b} are taken to be angles relative to some 
reference axis in the $xy$ plane, and the states are taken to be {\it invariant under 
rotations} about the $z$ axis. Similarly, in the present test, by choosing an experimental 
configuration for which the correlation function is $p_{12}({\bf a}, {\bf b}) = 
\cos^2(\vartheta_{ab})$, and, without loss of generality, by choosing for simplicity 
the alternative directions of the polarization analyzers so that $|{\bf a} - {\bf b}| = 
|{\bf a}'-  {\bf b}| = |{\bf a}'- {\bf b}'| = (1/3) |{\bf a} - {\bf b}'| = 
\vartheta_{ab}$ the inequality (\ref{eq1}) becomes:

\begin{equation}
S(\vartheta_{ab}) \equiv \frac{3p_{12}(\vartheta_{ab}) - p_{12}(3\vartheta_{ab})}{p_1(a') + p_2(b) } \le 1.
\label{eq2}
\end{equation}

\noindent where $\vartheta_{ab}$ is the relative angle between the polarizers' axes, 
$p_{12}({\bf a}, {\bf b}) = \cos^2(\vartheta_{ab})$, and $p_1$ and $p_2$ depend  on the 
linear analyzers' orientations because the two photons of a pair are polarized and 
arrive at the analyzers with the same definite polarization in the $x$ direction 
(we take {\bf a = x}).

\section{Experiment}
\begin{figure}[h]
\centering
\includegraphics[width=4.0in]{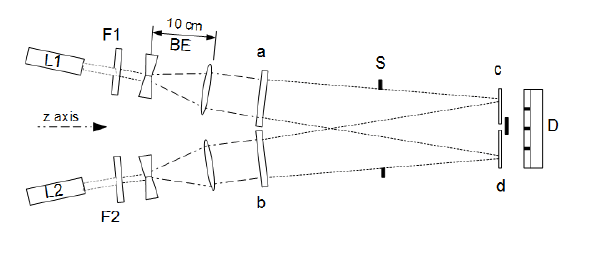}
\caption{Schematic drawing of the experiment, not on scale. Glancing incidence of beams: less than 2$^\circ$. 
           $L_1$ and $L_2$: lasers; $F_1$ and $F_2$: neutral density filters; BE: beam expander; 
           a, b, c, d: linear polarizers; S: screen; D: multianode detector. 
           Distance between BE and D: 2 m.}
\label{fig1}
\end{figure}

The conceptual scheme of the experiment is shown in the schematic drawing of Fig.\ref{fig1}.

Source A  emits coherent light with wave number k, source B emits coherent light 
with the same wave number k; the relative phase of the two sources is random; 
the light from each source propagates in the $z$ direction and it is unpolarized in 
the $xy$ plane. The linear polarizers a and b act as state-preparation devices, each  
converting the original state of the photon emitted from A and the photon emitted 
from B respectively into the preparation states $\Phi({\bf a})$ and $\Phi({\bf b})$, 
with definite polarization directions parallel to the polarizers' transmission axes 
represented by the unitary vectors {\bf a} and {\bf b} in the $xy$ plane. 

The light 
sources $L_1$ and $L_2$ are two unpolarized (632.8 nm) HeNe lasers, $1.0$ mm 
diameter, $\sim 1.5$ mrad divergence, output power from 
$L_1$ $\sim~1.0$ mW, output power from $L_2$ $\sim~3.0$ mW; and the two light  beams 
propagate at a glancing incidence smaller than $2^\circ$ along the $z$ direction. 
Neutral density filters reduce the rate of photons from each source to $\simeq 
1\cdot 10^3$ s$^{-1}$. Divergent lenses are used to increase  the size 
of each beam  up to $6$ cm in diameter, so producing their superposition upon  
the polarization analyzers c and d, each represented respectively 
by the unitary vector {\bf c} and {\bf d} in the $xy$ plane, and each respectively 
in front of detector \#1 and  detector \#2.  At $\lambda = 632.8$ nm, the polarizers' 
maximum  trasmission is 80\% and the extinction rate is $> 10^4$. 

The {\bf d} axis can be rotated about $z$ with respect to {\bf c} by the angle 
$\vartheta_{cd}$ in the range from $0^\circ$ to $360^\circ \pm 0.5^\circ$.  
Detector $D_1$ records only photons passing through c, and detector $D_2$ only 
photons passing through d.  Without lack of generality, we shall consider only 
the experimental configuration with each of the polarizers a, b, c oriented in 
the vertical direction  ({\bf a} = {\bf b} = {\bf c} = {\bf x}) and  the axis 
of polarizer d rotated by any angle $\vartheta = \vartheta_{cd}$ with respect 
to {\bf x}.

The coincidence measurements are based on the use of a position-sensitive 
detector and an advanced electronic acquisition system for coincidence recording. 
The detector is a Hamamatsu H8711 multianode photomultiplier (PMT) with $4\times 4$ 
anode pixels, pixel size $4.2$ mm $\times 4.2$ mm, quantum efficiency 0.15 at 
$\lambda = 632.8$ nm. The signal generated at each anode is first discriminated 
against the low-amplitude noise produced in the PMT, 
and then transmitted to the acquisition system. The photon arrival times at each 
pixel are recorded in a file for off-line data analysis. All the arrival times 
are measured by using an internal 40 MHz clock which provides 25 ns time 
resolution. The data acquisition system (DAQ) records counts during repeated  
cycles of 10 s each, with an  effective duty cycle of about 99.90\%, and the 
transfer process to the computer requires about 10 ms. 

\begin{figure}[h]
\centering
\includegraphics[width=4.0in]{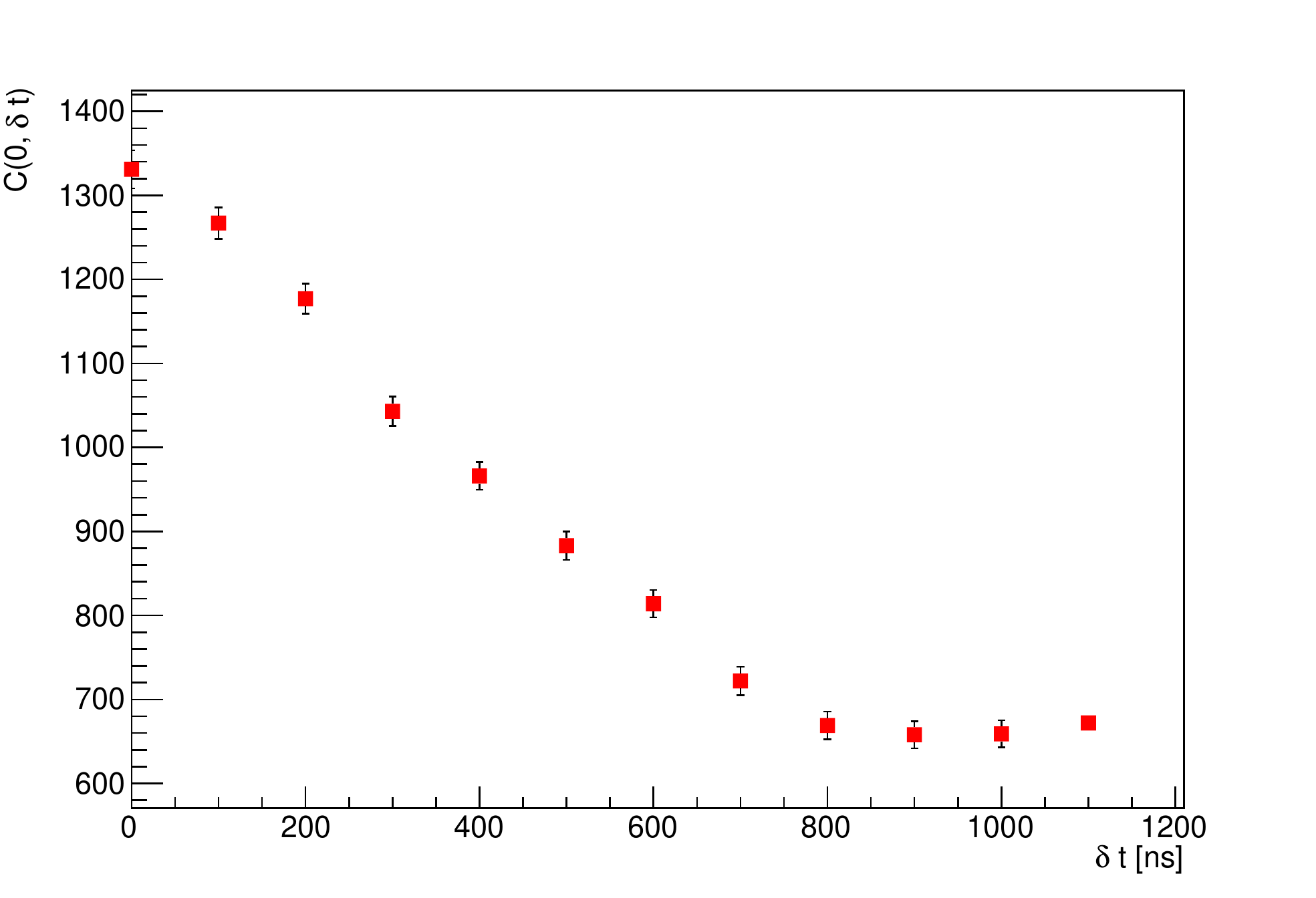}
\caption{$C(\vartheta=0^\circ,\delta t)$: number of experimentl coincidences as 
a function of the time delay $\delta t = t_2 - t_1$ with $t_2 > t_1$, at the angle 
$\vartheta = \vartheta_{cd} = 0^\circ$. 
Detectors: two 3-pixel vertical columns with relative horizontal separation $\simeq$ 
1.0 cm. The number of single counts at each detector was of the order of $\sim$
$10^6$. Full square: experimental points.}
\label{fig2}
\end{figure}

The multianode configuration of the photomultiplier allows to choose, during the 
off-line analysis, the two specific groupings of pixels forming the two distinct 
detectors $D_1$ and $D_2$, and generally each detector was formed of a vertical 
column of $2\div 4$ pixels. In the off-line analysis, the signals from 
such two detectors can be time-correlated by detecting their coincidence rate 
$C(\vartheta, \delta t)$ at any angle $\vartheta$ as a function of their 
relative time delay $\delta t$. The coincidence detection is based on the 
measurement of the arrival times $t_1$ and $t_2$ of the two signals generated 
respectively at a first pixel grouping forming detector $D_1$  
and at a second pixel grouping forming detector $D_2$. 
The number of coincidences, shown in Fig.\ref{fig2}, is obtained by counting 
all the pairs of photons recorded within a time delay $\delta t = t_2 - t_1$ 
with $t_2 > t_1$, and the resolution time of the coincidence recording is 
$\sim 100$ ns. 
The acquisition time of the measurement at each value of $\vartheta$ was in 
the range of 20-24 h, collecting data from 600 $\div$ 800 runs each of 100 s duration. 
Since we could analyze the data from the single runs, we have verifed that the stability 
of our system was a few $\%$.

Now, because the measuring procedure just described is based on the measurement 
of the arrival times of a pair of particles at two distinct detectors, let us recall 
the following fact. For chaotic light, both classical and quantum theories predict 
the same positive correlation in the distributions of photon coincidences. Such bunching 
effect, brought to light for the first time by Hanbury-Brown and Twiss(HBT) \cite{ref16} and 
clearly discussed, among others, in Refs. \cite{ref17,ref18}, can be observed {\it if}
the coincidences are recorded with a response time of the detecting system 
shorter than the time interval characteristic of the photon beam (e.g. its 
coherence time or the average time separation of the photons).

Thus, in the off-line analysis of our experiment, with parallel orientations 
of the linear polarizers c and d or in their absence, we have measured the 
coincidence counting as a function of their time delay $\delta t$ and we have 
observed that, as expected, in the range of $\delta t$ from 0 to a few hundreds 
of nanoseconds, there are twice as many coincidence counts as those recorded 
for larger $\delta t$.  

In Fig.\ref{fig2} we show a coincidence curve obtained with the axes of polarizers 
c and d oriented in the same direction {\bf c} = {\bf d} = {\bf x} ($\vartheta_{cd}=0^\circ$); 
at long delays $\delta t$, the coincidence probability is simply the classical 
joint probability of detecting the two uncorrelated beams at $D_1$ and $D_2$. 
A rotation of polarizer d will reduce the counts (at both shorter and longer delays)
as a function of the angle $\vartheta_{cd}$, because the fraction of all of these 
particles that pass through d, upon which they arrive polarized {\bf c} = {\bf x}, 
will decrease by the factor $\cos^2(\vartheta_{cd})$. 

Thus, if we normalize the counts recorded at different angles $\vartheta_{cd}$ 
with respect to the counts measured with polarizers c and d oriented in the same direction 
({\bf c} = {\bf d} = {\bf x}), since $p_1({\bf c}) = p_2({\bf d}) = 1$ for 
ideal instrumentation, the coincidence detection probability is simply
$\cos^2(\vartheta_{cd})$. Briefly, with a chosen disposition of polarizers 
a, b, c, d and for ideal instrumentation, the predicted joint probability of 
recording simultaneously two particles respectvely at $D_1$ and $D_2$ depends only  
on the relative angle $\vartheta_{cd}$:                                               

\begin{equation}
p_{12}({\bf c}, {\bf d}) = p_1({\bf c},1)\cdot p_2({\bf d},2 | 1) = \cos^2(\vartheta_{cd}).
\label{eq3}
\end{equation}
                                                                                                                        
Therefore, expression (\ref{eq3}) gives the polarization correlation relation to be 
substituted into the CH inequality to test its possible violations. Our measuraments 
have confirmed this predition, and we have used the experimental values of 
$\cos^2(\vartheta_{cd})$ to test inequality (\ref{eq2}).

\begin{figure}
\centering
\includegraphics[width=4.0in]{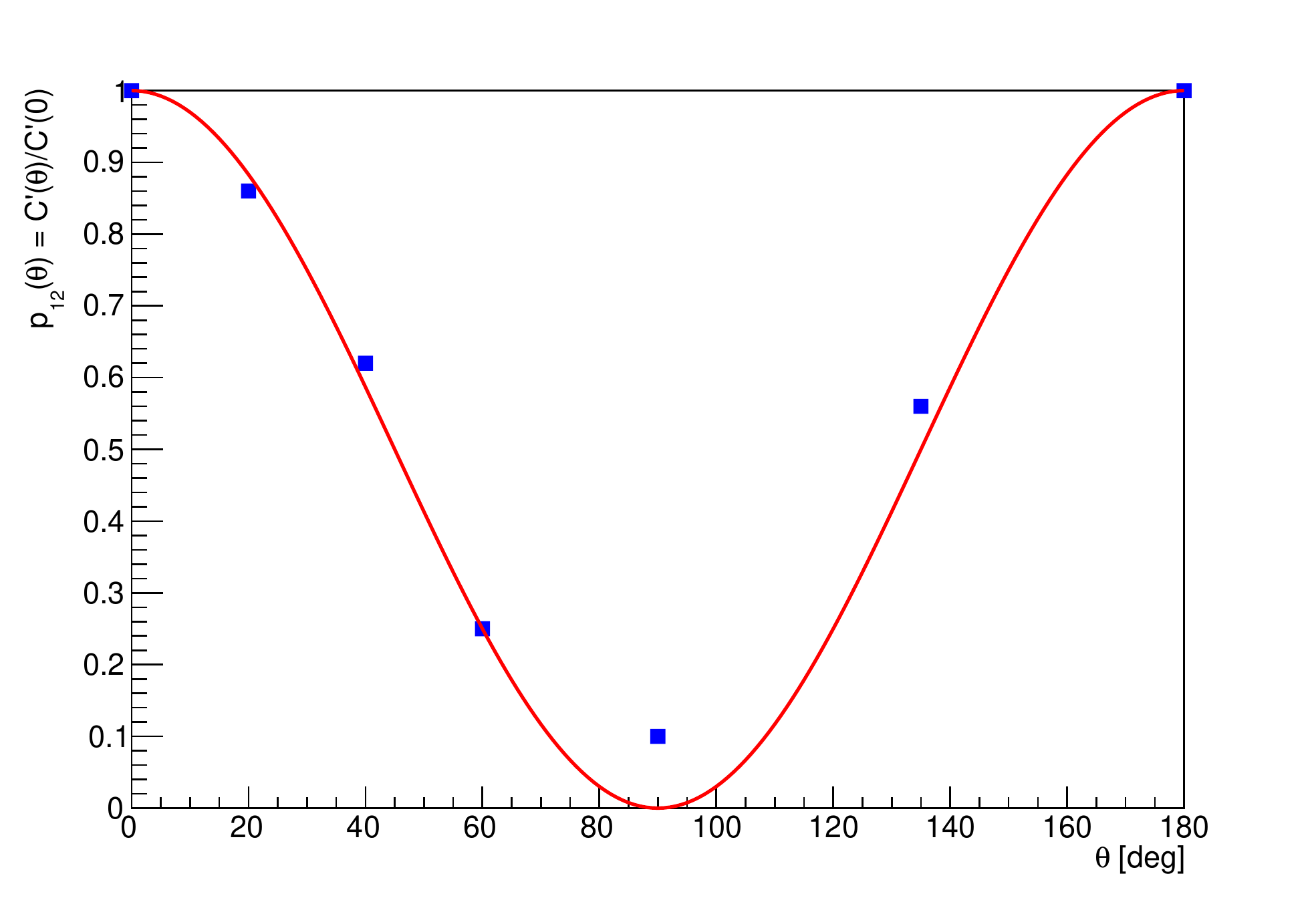}
\caption{$p_{12}(\vartheta) = C'(\vartheta)/C'(0)$, where $\vartheta=\vartheta_{cd}$ and 
$C'(\vartheta) = \sum_{\delta t} C(\vartheta, \delta t)$, at the same photon flux on the multianode.
Full square: experimental points after substraction of electronic and backround noise; 
error: $\simeq 0.2$ \%.  Full line: $\cos^2(\vartheta)$.}
\label{fig3}
\end{figure}

Fig.\ref{fig3} shows the outcome (after all, very predictable) of our experimental test of 
(\ref{eq3}). The acquisition time of the measurement at each value of $\vartheta$ was in 
the range of 20-24 h. 
Analogous result was obtained in an alternative test where the two beams from $L_1$ and 
$L_2$ remained separate from each other along their parallel paths to $D_1$ and $D_2$.

Notice also that, in experiments whose configuration is such that {\bf a} $\ne$ {\bf b},  
by normalizing the coincidence counting with respect to the coincidences obtained with  
{\bf a} = {\bf c} and  {\bf b} = {\bf d}, $p_{12}({\bf c}, {\bf d})$ is still given by 
a simple trigonometric function:

\begin{equation}
p_{12}({\bf c}, {\bf d}) = p_1({\bf c}, 1)\cdot p_2({\bf d}, 2 | 1) = 
\cos^2( \vartheta_{cd} - \vartheta_{ab})
\label{eq4}
\end{equation}

\noindent with $\vartheta_{cd} \ge \vartheta_{ab} = \vartheta_{cb}$.

\begin{figure}
\centering
\includegraphics[width=4.0in]{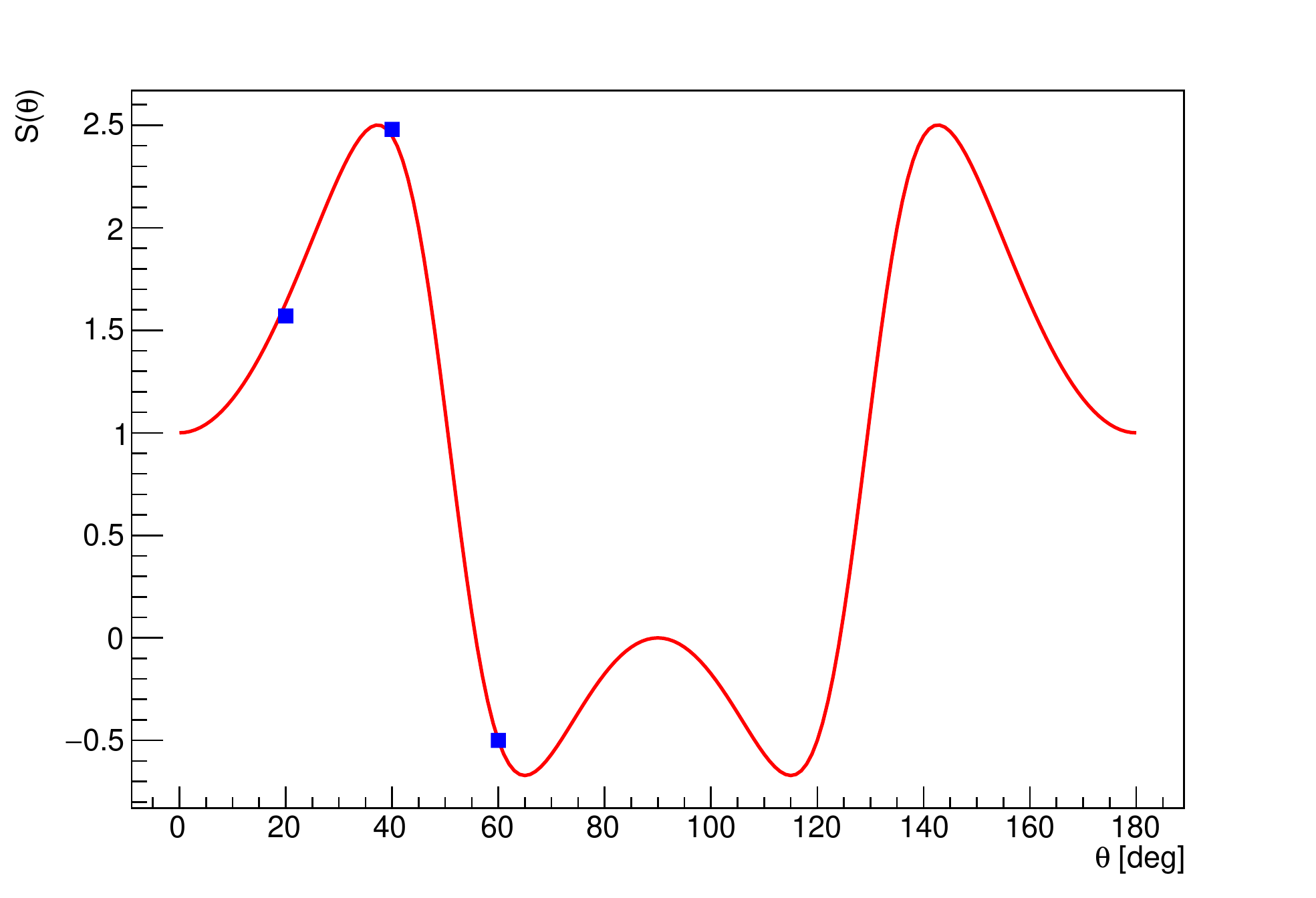}
\caption{Full line : ideal value of $S(\vartheta)$ calculated for the present experiment, 
showing that violation of inequality (\ref{eq2}) is predicted for a wide range of 
$\vartheta=\vartheta_{cd}$. Full square: experimental values measured at $\vartheta=20^\circ,~
40^\circ$ and $60^\circ$.}
\label{fig4}
\end{figure}

Now, if we substituite into $S(\vartheta)$ the predictions calculated for our experiment, 
$p_1({\bf a}') = p_1(2\vartheta) = \cos^2(2\vartheta),~p_2({\bf b}) = p_2(\vartheta) = 
\cos^2(\vartheta),~p_{12}(\vartheta) = cos^2(\vartheta)$, we obtain the ideal value of 
$S(\vartheta)$ shown in Fig.\ref{fig4} which violates the CH inequality whenever it 
exceeds 1. 
We have  experimentally confirmed such violations by substituiting into $S(\vartheta)$ 
the measured values of the probabilities. 

In particular we have obtained:
     
\begin{equation}
S(20^\circ) = 1.57 \pm 0.02, ~~~~~ S(40^\circ) =  2.48 \pm 0.03
~~~~~ {\rm and} ~~~~~ S(60^\circ) = -0.5 \pm 0.03
\label{eq5}
\end{equation}

We wish to stress that the measurement procedure of this experiment fulfilled 
all the conditions necessary to guarantee relative separation and independence 
of the two members of the pairs whose coincidences have been recorded (see also 
the introduction). 
Actually: 
\begin{itemize}
\item we have used two independent lasers, two independent optical paths, 
two independent photon detectors; 
\item we could, without observing whatever effect on one another, switch 
off separately and independetly each of the two lasers, or obscure each 
of the two optical paths, or switch off each of the two detectors; 
\item we have used two independent lasers in order to record mainly 
random coincidences between the two sources, and we have tested that 
the number of coincidences from one source (with the other source 
switched off) is of the same order of the coincidence background. 
\end{itemize}
Basically, we have simply tested the quantum-mechanical
equivalent of the classical Malus' law on an incoherent beam of
polarized single photons.

\section{Discussion}
In the introduction we have recalled that, in Bell-type experiments performed 
with pairs of polarization-correlated photons, the simple sinusoidal 
correlation relation $p_{12}({\bf a}, {\bf b}) = 1/2\cos^2(\vartheta)$ 
leads to violations of Bell's theorem, and in particular of the CH inequality. 
Since Bell's theorem has been derived with no use of quantum mechanics,
we have wondered if any physical systems might 
be found, formed of pairs of {\it not-entangled photons emitted into the experimental 
apparatus from two independent sources}, that nevertheless might experimentally exhibit 
(at the detectors) similar sinusoidal relations, with consequent violations of the 
CH inequality. Actually the present experiment shows that such a test is feasible, 
and its outcome has confirmed our reflection.

Now we wish to suggest our interpretation of this result. Let us first recall that the 
state-preparation of a physical system is not a form of measurement.
Following L.E. Ballentine \cite{ref19}: "...The essential 
distinction between the two concepts are that state-preparation refers to the future, 
whereas measurement refers to the past; and equally important, that measurement involves 
{\it detection} of a particular system, whereas state-preparation provides conditional 
information about a system {\it if} it passes through the apparatus". The concepts of 
state preparation and measurement occasionally coincide, and then the experimental 
outcome will directly describe the state-preparation of the photon pair.

In optical polarization experiments, the relative rotation of the polarizers' 
axes is a procedure of state-preparation that gives conditional information on the relative 
polarization of the photons, {\it if} they the pass through the polarizers; the measurement being 
on the contrary performed by the detectors and coincidence counter. The measured values of 
the coincidence probability $p_{12}({\bf a}, {\bf b})$ will simply mirror the final 
preparation-states of the pairs (at the detectors, after their passage through the 
polarizers) as a function of {\it the experimental disposition of the apparatus}, in 
particular as a function of the setting of polarizer\#2 (at the crossing time of 
particle\#2) relative to the setting of the spatially-separated analyzer\#1 
(at the crossing time of particle\#1). 
In other words, our measurements of the coincidence probability $p_{12}(\vartheta)$ as a
function of $\vartheta$ simply reveal in what final state the two {\it uncorrelated} members 
of a pair with definite relative polarization orientations have been prepared; 
and actually this information is the one used in all the experiments performed to 
test Bell-like inequalities.

We may therefore understand that the measurement precedure adopted in 
the Bell-type experimets yields the polarization relation between the two members of a pair, 
either entangled or not entangeld, in their final preparation state.
In particular, for either quantum or classical physical systems, sinusoidal correlations 
relation of $p_{12}({\bf a}, {\bf b})$ substituited into the CH inequality, may cause its violation.
Consequently we believe that the CH inequality will be conclusively tested only by relating it 
to Bell-type experiments with an analysing instrument and a final state-preparation instrument 
that do not coincide.

\end{document}